\documentclass[iop]{emulateapj}
\usepackage{epsfig}

\begin{document}

	\title{Discovery of an unidentified Fermi object as a black widow-like millisecond pulsar}
	\author{A.~K.~H.~Kong\altaffilmark{1,10}, R.~H.~H. Huang\altaffilmark{1}, K.~S.~Cheng\altaffilmark{2}, J. Takata\altaffilmark{2}, Y. Yatsu\altaffilmark{3}, C.~C.~Cheung\altaffilmark{4}, D. Donato\altaffilmark{5,6}, L.~C.~C.~Lin\altaffilmark{7}, J.~Kataoka\altaffilmark{8}, Y. Takahashi\altaffilmark{8}, K.~Maeda\altaffilmark{8}, C.~Y.~Hui\altaffilmark{9}, P.~H.~T.~Tam\altaffilmark{1}}
	\altaffiltext{1}{Institute of Astronomy and Department of Physics, National Tsing Hua University, Hsinchu 30013, Taiwan; akong@phys.nthu.edu.tw}
	\altaffiltext{2}{Department of Physics, The University of Hong Kong, Hong Kong}
	\altaffiltext{3}{Department of Physics, Tokyo Institute of Technology
2-12-1, Ookayama, Meguro, Tokyo, 152-8551, Japan}
	\altaffiltext{4}{National Research Council Research Associate, National Academy of Sciences, Washington, DC 20001, resident at Naval Research Laboratory, Washington, DC 20375, USA}
	\altaffiltext{5}{CRESST and Astroparticle Physics Laboratory NASA/GSFC, Greenbelt, MD 20771, USA}
	\altaffiltext{6}{Department of Astronomy, University of Maryland, College Park, MD 20742, USA}
		\altaffiltext{7}{General Education Center, China Medical University, Taichung 40402, Taiwan}
	\altaffiltext{8}{Research Institute for Science and Engineering, Waseda University, 3-4-1 Okubo, Shinjuku, Tokyo 169-8555, Japan}
	\altaffiltext{9}{Department of Astronomy and Space Science, Chungnam National University, Daejeon, Republic of Korea}
	\altaffiltext{10}{Golden Jade Fellow of Kenda Foundation, Taiwan}

\newcommand{\chandra}{{\it Chandra}}
\newcommand{\asca}{{\it ASCA}}
\newcommand{\rosat}{{\it ROSAT}}
\newcommand{\xmm}{{\it XMM-Newton}}
\newcommand{\swift}{{\it Swift}}
\newcommand{\fermi}{{\it Fermi}}
\newcommand{\ufo}{1FGL\,J2339.7--0531}
\newcommand{\nufo}{2FGL\,J2339.6--0532}
\newcommand{\lum}{\thinspace\hbox{$\hbox{ergs}\thinspace\hbox{s}^{-1}$}}
\newcommand{\flux}{\thinspace\hbox{$\hbox{ergs}\thinspace\hbox{cm}^{-2}\thinspace\hbox{s}^{-1}$}}

\def\spose#1{\hbox to 0pt{#1\hss}}
\def\laeq{\mathrel{\spose{\lower 3pt\hbox{$\mathchar"218$}}
     \raise 2.0pt\hbox{$\mathchar"13C$}}}
\def\gaeq{\mathrel{\spose{\lower 3pt\hbox{$\mathchar"218$}}
     \raise 2.0pt\hbox{$\mathchar"13E$}}}

\begin{abstract}
The {\it Fermi Gamma-ray Space Telescope} has revolutionized our
knowledge of the gamma-ray pulsar population, leading
to the discovery of almost 100 gamma-ray pulsars and
dozens of gamma-ray millisecond pulsars (MSPs). Although
the outer-gap model predicts different sites of emission
for the radio and gamma-ray pulsars, until now all of the known
gamma-ray MSPs have been visible in the radio.
Here we report the discovery of a ``radio-quiet'' gamma-ray emitting MSP candidate by using \fermi, \chandra, \swift, and optical observations. The X-ray and gamma-ray properties of the source are consistent with known gamma-ray pulsars. We also found a 4.63-hr orbital period in optical and X-ray data. We suggest that the source is a black widow-like MSP with a $\sim 0.1M_\odot$ late-type companion star. Based on the profile of the optical and X-ray light-curves, the companion star is believed to be heated by the pulsar while the X-ray emissions  originate from pulsar magnetosphere and/or from intra-binary shock. No radio detection of the source has been reported yet and although no gamma-ray/radio pulsation has been found, we estimated that the spin period of the MSP is $\sim 3-5$ ms based on the inferred gamma-ray luminosity.

\end{abstract}


\section{Introduction}
The \fermi\ Large Area Telescope (LAT) has detected 1873 $\gamma$-ray point sources during its first 24 months of operation (Abdo et al. 2011). The majority of the extra-galactic sources have been identified 
as Active Galactic Nuclei (AGN; Abdo et al. 2010a), while many of the Galactic sources 
has been identified as $\gamma$-ray emitting pulsars (Abdo et al. 2010b). 
Gamma-ray pulsars are typically either young and
energetic, such as the Crab and Vela pulsars, or
very short period millisecond pulsars (MSPs). Throughout
this paper we will refer to these two classes as
the energetic and millisecond pulsars.
In this second source catalog (2FGL), there are more than hundred identified energetic and millisecond pulsars. About half of them with gamma-ray pulsations were discovered previously by using radio pulsation search (e.g. Ransom et al. 2011; Keith et al. 2011; Cognard et al. 2011; Caraveo 2010). 
In addition, although there were 35 gamma-ray pulsars
identified from the energetic class using a blind fourier
search of the LAT data (see Pletsch et al. 2011 and references therein), only four were subsequently
identified as radio pulsars by folding the radio data with the pulsar ephemerides (Camilo et al. 2009; Saz Parkinson et al. 2010; Pletsch et al. 2011).
This indicates that there is a population of ``radio-quiet'' energetic pulsars as seen from the Earth. However, all 21 gamma-ray emitting MSPs are ``radio-loud''.
Based on current observations, we can classify 
the current population of $\gamma$-ray emitting pulsars into ``radio-loud'' energetic pulsars,  ``radio-quiet'' 
pulsars and ``radio-loud'' MSPs. However, it is unclear if
 there is a class of ``radio-quiet'' $\gamma$-ray emitting MSPs. 

It is well known that the radio emissions of pulsars are associated with the 
activity of polar cap accelerator.
  On the other hand,  
the high-energy emissions from the pulsar magnetosphere have been studied
with  polar cap  model
(Ruderman \& Sutherland  1975),
 slot gap model (Muslimov \& Harding  2003) and outer gap model
 (Takata, Wang \& Cheng 2010a). 
The polar cap model assumes the emission region is close to the stellar surface
and above the polar cap,  and therefore the model 
 implies radio-loud $\gamma$-ray 
pulsars are much more common than radio-quiet $\gamma$-ray pulsars.  
The outer gap and slot gap models
 assume an acceleration region extending to the outer magnetosphere and 
indicates radio-quiet $\gamma$-ray pulsars are the major population 
of $\gamma$-ray pulsars.  Based on the spectral shape in the GeV bands and the population of radio-quiet $\gamma$-ray pulsars found with \fermi\ (Abdo et al. 2010b), 
the outer gap or slot gap region is more favorable as the origin of the 
$\gamma$-rays from energetic pulsars (Takata, Wang \& Cheng 2010). 

For MSPs, ``radio-quiet'' millisecond $\gamma$-ray 
pulsars have not been identified so far. This may be due to following reasons: 
(1) the present sensitivity of blind frequency search prevents a detection  
of millisecond pulsation in \fermi\ data; (2) most of known radio MSPs are in binaries that gamma-ray searches would be insensitive to, and (3) $\gamma$-ray emissions from MSPs always accompany radio emissions.
On the other hand, Venter, Harding \& Guillemot (2009) 
found that the pulse profiles  of  six MSPs 
detected by the \fermi\ can be fitted by 
 the geometries with outer gap or the slot models. 
This implies that the emission regions of the $\gamma$-rays are different from 
the radio emission site for some of the MSPs and thus we expect that there are ``radio-quiet'' $\gamma$-ray MSPs for which the radio beam is outside the line of sight. It is worth noting that ``radio-quiet'' here does not mean no intrinsic radio emission, just none beamed in our direction.

In this Letter, we report a multiwavelength identification of a ``radio-quiet'' $\gamma$-ray MSP candidate. In a separate paper, Romani and Shaw (2011) arrived at similar conclusions via optical spectroscopy.

\section{Unidentified Fermi source as a ``radio-quiet'' $\gamma$-ray emitting millisecond pulsar}

To identify suitable targets of ``radio-quiet'' $\gamma$-ray MSPs, we first selected candidates from the \fermi\ LAT first source catalog (1FGL; Abdo et al. 2010c) based on four criteria: 1) No known association at other wavelengths; 2) source variability; 3) Galactic latitude, and 4) gamma-ray spectral shape. 

We used the variability index in the 1FGL catalog to characterize source variability. Gamma-ray pulsars have always been found to be
steady sources of gamma-ray emission (Abdo et al. 2010c,2011).
This property can therefore be used to help identify
which of the unidentified sources are probably pulsars. The 1FGL catalog defines a variability index, for which a value greater than 23.21 means that there is less than a 1\% probability of being a steady source. We therefore selected sources with a variability index less than 23. 

MSPs are in general older than energetic $\gamma$-ray pulsars. Young objects like energetic $\gamma$-ray pulsars ($\tau < 1~$Myr) are likely located in the Galactic plane, while a fraction of MSPs should be at higher Galactic latitudes. We thus selected $Fermi$ sources with high Galactic latitudes ($\left|b\right| >40^{\circ}$). The choice of this value is based on Monte-Carlo simulations for the Galactic population of MSPs, which show that no energetic $\gamma$-ray pulsar exists above $\left|b\right|=40^{\circ}$ (Takata, Wang \& Cheng 2011). 

Finally, we identified potential candidates from the $\gamma$-ray spectra. Although only a power-law spectrum is listed in the first-year catalog, it also has a curvature index to indicate how good of a power-law fit. For $\gamma$-ray pulsars, their $\gamma$-ray spectra is usually described by a power-law plus an exponential cutoff model (Abdo et al. 2010b). According to 
 the 1FGL catalog, the curvature index, for which a value greater 
than $C=11.34$ indicates less than 1\% chance that the power-law spectrum 
is a good fit. Hence we chose objects with a curvature index larger than 12.

If a 1FGL source satisfied all four criteria, we short-listed as a potential candidate. From the short list, we further searched for X-ray imaging data ({\it Chandra}, {\it XMM-Newton} and {\it Swift}) from public archive and looked for X-ray sources within the error circles of \fermi\ as the first step for multiwavelength investigation. In this paper, we focus on one of our targets, \ufo\footnote{We will use the 1FGL naming conversion throughout the paper, and the source is called 0FGL\,J2339.8-0530 and \nufo\ in the bright source catalog and 2FGL catalog, respectively.}.

\section{Multiwavelength Identification}

\ufo\ is one of the 205 bright gamma-ray sources detected with \fermi\ LAT during its first three months of operation (Abdo et al. 2009). It remains as an unidentified source in the second \fermi-LAT source (2FGL) catalog. In the 1FGL catalog (Abdo et al. 2010c), \ufo\ has a variability index of 9.2 and a curvature index of 22.7. By comparing the 1FGL and 2FGL catalog (Abdo et al. 2011), the gamma-ray flux of \ufo\ is constant and near $3\times 10^{-11}$ ergs  s$^{-1}$ cm$^{-2}$. With better statistics, the second source catalog shows that a log parabola spectrum can provide a better spectral fit comparing to a power-law model (Abdo et al. 2011). With its high Galactic latitude ($-62^{\circ}$) and unidentified nature in all \fermi\ catalogs, \ufo\ is therefore one of the best targets to search for ``radio-quiet'' $\gamma$-ray MSPs.

\ufo\ was observed with \chandra\ and \swift/XRT for 21ks on 2009 October 13 (PI: Cheung), and 3.2ks on 2009 November 4, respectively. The \chandra\ imaging was operated with ACIS-I while the XRT was in the photon counting imaging mode. We reprocessed with updated calibration files. HEASOFT version 6 and CIAO version 4.2 were used for data reduction and analysis. Within the 95\% error circle of \fermi\ (based on the refined position in the 2FGL catalog), there is only one relatively bright X-ray source in both observations. The brightest X-ray source (CXOU J233938.7-053305) near the center of the error circle has an X-ray flux of $3\times 10^{-13}$ ergs s$^{-1}$ cm$^{-2}$ (0.3--10 keV) based on a spectral fitting using an absorbed power-law with a best-fit photon index of 1.1 (based on the \chandra\ observation). This indicates that the ratio between the X-ray and $\gamma$-ray ($> 100$ MeV) flux ($F_{\gamma}\sim 3\times 10^{-11}$ ergs s$^{-1}$ cm$^{-2}$) as measured by \fermi\ (Abdo et al. 2011) become $F_{X}/F_{\gamma}\sim 0.01$, which 
is consistent with typical observed values for $\gamma$-ray pulsars. It is worth noting that there are 9 much fainter \chandra\ sources in the error circle and their X-ray-to-gamma-ray flux ratios are less than 0.1\%. Although we cannot totally rule out their association with the gamma-ray source, such a low flux ratio is not typical. We therefore identified CXOU J233938.7-053305 as the potential X-ray counterpart to \ufo.

Within the \chandra\ error circle (0.6 arcsec at 90\% level) of CXOU J233938.7--053305, there is a $R\sim19$ star from the USNO catalog (Monet et al. 2003). The same optical source is also seen in the SDSS Data Release 8 images as SDSS\,J233938.74-053305.2 with $u'=20.85$, $g'=19.0$, $r'=18.61$, $i'=18.25$, and $z'=18.23$, as well as in the GALEX images with NUV=22.88 (177--283 nm). \swift/UVOT observations taken simultaneously with the X-ray observations detected the source with $U=19.58$, $B=19.53$, and $V=18.88$.
The X-ray-to-optical flux ratio (390) is much too high for a foreground star and an AGN (e.g. Green et al. 2004, Laird et al. 2009). Moreover, the spectral energy distribution from UV to optical indicates that it is unlikely to be an AGN (e.g. Richards et al. 2002). Instead, it looks like a F- or G-type star. In a separate paper (Romani \& Shaw 2011), they show via optical spectroscopy that the stellar object is consistent with a late-type star. We also had a long-term optical monitoring program with the MITSuME 50cm telescope (Kotani et al. 2010) located at Akeno Observatory in Japan from 2010 September to 2010 November. MITSuME equips with a tricolor camera that can perform simultaneous imaging in the $g'$, $R$, and $I$ bands.
It is clear from Table 1 that all three bands show noticeable variability.
Based on the above observations, we believe that the gamma-ray/X-ray source is likely a binary system with optical emission from the irradiation of the companion star, while the central compact object is responsible to the X-ray and gamma-ray emission.

\begin{table*}
\begin{center}
\footnotesize
\caption{Optical Observation log of \ufo}
\begin{tabular}{lccc}
\hline
\hline
Date and Time (UT) & Telescope & Filter & Exposure$^a$ \\
\hline
2010-10-25 14:44--15:57 & Lulin 1m & $V$ & 2 min\\
2010-10-26 13:33--15:59 & Lulin 1m & $V$ & 2 min\\
2010-10-31 10:23--15:53 & Lulin 1m & White & 2 min\\
2010-11-01 11:33--16:27 & Lulin 1m & White & 2 min\\
2010-11-11 01:45--06:20 & Tenagra 0.8m & White & 5 min\\
\\
2007-09-13 22:32:08 & GALEX & NUV=22.88 & 1642 sec\\
2008-11-20 05:25--05:29 & SDSS & $u'=20.85$, $g'=19.0$, $r'=18.61$, $i'=18.25$, $z'=18.23$ & 54 sec\\
2009-11-04 07:56--16:07 & UVOT & $U=19.58$, $B=19.53$, $V=18.88$ & 1144, 838, 1144 sec\\
2010-09-13 14:52:26 & MITSuME 0.5m & $g'=18.51$, $R=18.04$, $I=17.53$ & 60 sec$^b$\\
2010-09-17 14:46:37 & MITSuME 0.5m & $g'=18.54$, $R=17.98$, $I=17.63$ & \\
2010-09-20 13:17:30 & MITSuME 0.5m & $g'=18.15$, $R=17.33$, $I=17.38$ & \\
2010-10-04 15:55:20 & MITSuME 0.5m & $g'=18.51$, $R=17.93$, $I=17.53$ & \\
2010-10-10 14:28:06 & MITSeME 0.5m & $g'=19.12$, $R=18.53$, $I=17.94$ & \\
2010-11-09 13:21:22 & MITSeME 0.5m & $g'=18.43$, $R=17.82$, $I=17.21$ & \\
2010-11-10 13:22:43 & MITSeME 0.5m & $g'=18.98$, $R=18.12$, $I=17.51$ & \\
2010-11-11 13:19:40 & MITSeME 0.5m & $g'>20.19$, $R=19.44$, $I=19.04$ & \\
2010-11-12 11:22:28 & MITSeME 0.5m & $g'=18.43$, $R=17.70$, $I=17.44$ & \\
2010-11-14 12:39:55 & MITSeME 0.5m & $g'=19.00$, $R=18.16$, $I=17.84$ & \\
2010-11-15 12:40:11 & MITSeME 0.5m & $g'=18.07$, $R=17.91$, $I=17.51$ & \\
2010-11-18 12:53:05 & MITSeME 0.5m & $g'>20.02$, $R=21.77$, $I=19.16$ & \\
2010-11-21 12:46:57 & MITSeME 0.5m & $g'=18.70$, $R=17.61$, $I=17.49$ & \\
\hline \hline
\end{tabular}
\end{center}
\par
\smallskip
\hspace{1.2cm}	$^a$ Exposure time for each frame.\\
\hspace*{1.2cm}	$^b$ Some frames were combined to increase the signal-to-noise ratio.\\
\end{table*}

Because \ufo/CXOU J233938.7--053305 may be a new class of interesting object, we carried out an intensive optical monitoring campaign using the 1m telescope at the Lulin Observatory in Taiwan and the 0.8m Tenagra Telescope in Arizona (see Table 1 for the observation log). All images are flat-field and bias corrected and we performed relative photometry by comparing with several comparison stars in the field. In all 5 nights, the light-curves clearly show variability on a timescale of 4--5 hours (see Figure 1 \& 2). 
We then performed a timing analysis by using the Lomb-Scargle periodogram on 
the combined optical data and found a period of 4.6342(9) hours. 
Figure 2 shows the folded light-curves of all the optical data (upper panel) 
and \chandra\ data (lower panel) at a period of 4.6342 hours. The phase zero is 
chosen at MJD 55500. In addition, we folded the optical colors ($g'-I$) obtained with MITSuME and X-ray hardness ratio (1.5--8keV/0.3--1.5keV) at the same period (Figure 2).

\begin{figure}
	\centering
	\psfig{file=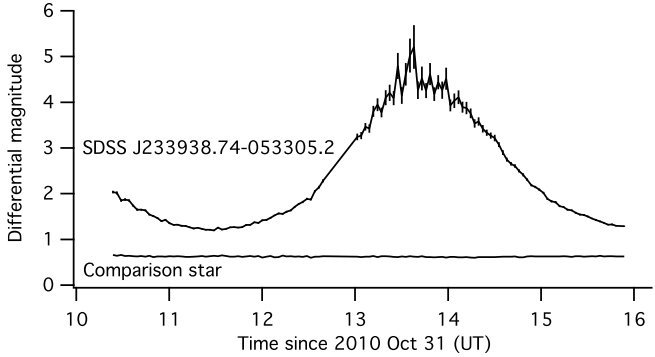,width=3.65in}
	\caption{Representative optical (white light) light-curve of the optical counterpart to \ufo, as observed by the Lulin 1m telescope in Taiwan. The differential magnitudes are derived by comparison with several comparison stars in the field. Periodicity on a timescale of 4--5 hours is clearly seen. Also shown is the light-curve of a comparison star (with an average error of 0.01--0.02 magnitudes). The timing resolution is 2 min.}
\end{figure}

\begin{figure*}
	\centering
	\psfig{file=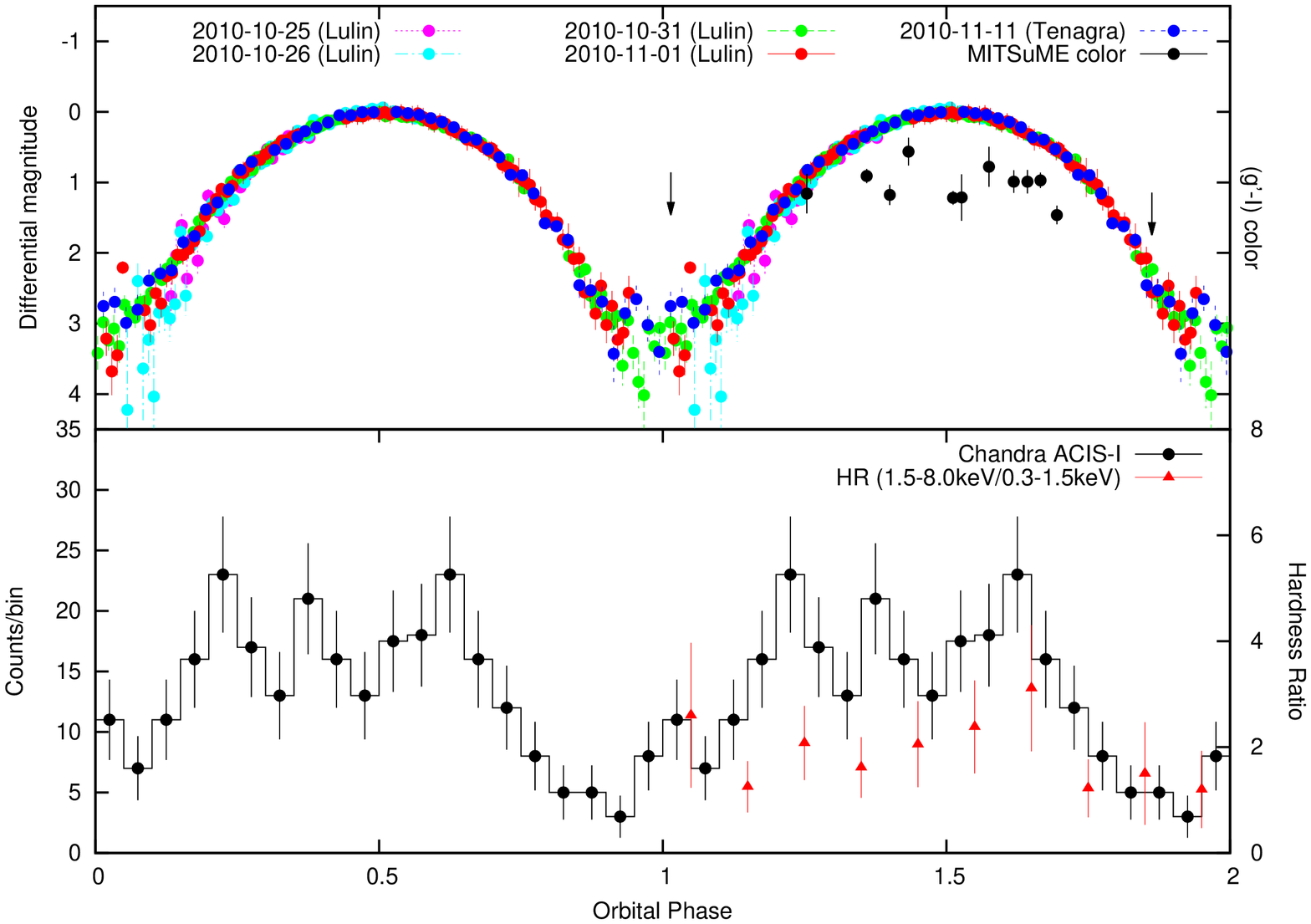,width=4.6in,angle=-90}
	\caption{Folded light-curve of optical and \chandra\ observations of \ufo\ with a best-fit period of 4.6342 hr. Optical colors ($g'-I$) obtained with MITSuME are also plotted. The phase zero is defined as 2010 October 31 (MJD 55500). Also plotted with the X-ray light-curve is the X-ray hardness ratio (1.5--8 keV/0.3--1.5 keV) with triangles. It is evident that both optical and X-ray light-curve show similar modulation. The X-ray hardness ratio exhibits some variability when the X-ray light-curve is at its minimum. Note that the \chandra\ data only cover about one orbital period.}
\end{figure*}

\section{Discussion}

Using multi-wavelength data, we found an X-ray/optical counterpart to \ufo. The X-ray-to-gamma-ray flux ratio is consistent with a gamma-ray pulsar while the gamma-ray spectrum deviates from a power-law which is commonly seen in MSPs. The X-ray spectrum is very hard with a power-law photon index of 1.1.  Such a hard spectrum is not typical for X-ray binary, AGN, and foreground star. However, some MSPs show similar spectral characteristics (e.g., Archibald et al. 2010; Tam et al. 2010) and such a hard spectrum could be from a pulsar wind nebula.
We propose that \ufo\ is a gamma-ray emitting MSP in a binary system and the optical emission is from a late-type companion star with contribution from the heating on the stellar surface. 

The optical light-curve shows clear modulation at a period of 4.6342 hours (Figure 2). It is beyond any doubt that the periodicity is associated with the orbital period of the system. Given the large variation (about 3--4 magnitudes) in the optical light-curve, the inclination must be high. Better photometric data in the near infrared and light-curve modeling will be able to constrain the inclination in the future. Assuming the companion has filled its the Roche lobe, the mass limit of the companion star can be estimated from the orbital period. For a 4.6-hr orbital period, the mass for a normal main-sequence star is $< 0.5 M_\odot$ (Wilson et al. 1999).

The large optical variability and hints of color variation (Fig. 2) suggest that the companion is being heated by the pulsar. 
The irradiation of pulsar $\gamma$-ray emissions onto the companion star can produce the orbital modulation of the optical emission from the binary system (e.g. Takata, Cheng \& Taam 2010, 2011). In this model, the optical maximum occurs at the inferior conjunction, where the pulsar is located
between the companion star and the Earth. Furthermore, during the optical maximum, the color of the star tends to be bluer because of the heating effect. There is some indication from the MITSuME data that the color of the star has become bluer during the optical maximum (Fig. 2). 
The maximum luminosity of the optical emission can be
estimated as $\delta L_{opt}\sim (\pi \theta^2/\delta\Omega) L_{\gamma}\sim 10^{31} (\theta/0.1 \mathrm{rad})^2(\delta\Omega/3 \mathrm{rad})^{-1}(L_{\gamma}/10^{33}$ ergs s$^{-1}$) ergs s$^{-1}$, where $\theta$ is the angular size of the
companion star measured from the pulsar and $\delta\Omega$ is the solid angle of the $\gamma$-ray beam. The observed maximum $R$-band magnitude of \ufo\ is about 18 (see Table 1) which is consistent with the above estimation.

It has been suggested that X-ray emissions from black widow systems originate from pulsar
magnetosphere and/or from intra-binary shock due to the interaction between the pulsar wind and the
injected material from the low mass star (e.g., Kulkarni et al. 1992, Stappers et al. 2003). If the X-rays are produced in the magnetosphere, the orbital
modulation will be produced due to the physical eclipse of the pulsar by the companion star. In this case, the minima of the X-ray and optical light-curves should be at the same phase, and we indeed observed that (see Fig. 2). Alternatively, 
if the X-rays originate from the intra-binary shock region, an effect of the Doppler boosting relating with the post-shock flow can produce the orbital modulation (Arons \& Tavani 1993). It has been suggested for the pulsar
and massive stellar system (e.g. PSR B1259-63/LS 2883 system, Bogovalov et al. 2008), the post-shock
flow can be accelerated into relativistic regime because of an explosion in the downstream region (see also Tam et al. 2011). The Doppler boosting increases the intensity as $I_{\nu}\propto D^{3+\alpha} I_{\nu}^{'}$, where $I_{\nu}
^{'}$ is the intensity in the comoving frame, $D$ is Doppler factor and $\alpha$ is the spectral index
in the co-moving frame. An orbital modulation by a factor of five may suggest the Doppler factor $D\sim 1.6$ with $\alpha\sim 0.5$. 

The X-ray hardness ratio roughly correlates with the X-ray light-curve (Figure 2). 
It may suggest that the pulsar (which has a very hard
X-ray spectrum) is physically eclipsed by the companion
and that the soft emission is an extra source of light (e.g. a wind shock).
There is some support for this in the assymetry of the optical light curve.
Much better X-ray and optical photometric data are required for serious modeling.

The source has a very low absorption ($3\times10^{20}$ cm$^{-2}$) along the line of sight. In comparison with Geminga with $N_H =1.5\times10^{20}$ cm$^{-2}$ and a distance of 150--250 pc, and SAX J1808.4--3658 with $N_H=4\times10^{20}$ cm$^{-2}$ and a distance of 1.3 kpc. The distance of \ufo\ is therefore likely to be between 300 and 1000 pc. If the companion star is a late-type main sequence (e.g., a M5 dwarf), and has $V\sim21$ during optical minimum when the heating effect is at minimum, the implied distance is about 700 pc.

If we assume a distance of 700 pc, then the gamma-ray power is
$L_{\gamma} \sim 10^{33}$ ergs s$^{-1}$.
Most of MSPs have an efficiency about 10\% which gives spin-down luminosity
$L_{sd} \sim 10^{34}$ ergs s$^{-1}$. According to Takata, Cheng \& Taam (2011), the spin period of the MSP can be derived from the spin-down power. Using the estimated spin-down power, the spin period will be $\sim 4$ ms. We note that the distance to the source is uncertain; for instance, Romani and Shaw (2011) estimated a distance of $1.1\pm0.3$ kpc. If we assume a distance error of a factor of 2 (i.e., 350--1400 pc), the spin period will be 3--5 ms.  

The bright \fermi\ source \ufo\ is an intriguing object that is potentially 
 the first ``radio-quiet'' $\gamma$-ray emitting MSP in a binary system. 
 Despite \ufo\ being one of the brightest unidentified pulsar-like \fermi\ sources located at high Galactic latitude, no radio detection has been reported. Based on a recent pulsation search with the Green Bank Telescope, \ufo\ was not detected (Ransom et al. 2011). It may indicate that \ufo\ is ``radio-quiet'', or the radio emission must be very weak.
 
The system resembles black widow MSP for which the MSP has a very low-mass ($< < 0.1M_\odot$) companion in an orbit less than a day. The optical light-curves of black widow systems usually shows large orbital variation because the irradiation produces strong heating on the companion facing the pulsar. It is therefore suggestive that the companion (usually a white dwarf) is being evaporated by the high-energy radiation from the pulsar. On the other hand, there is a class of black widow MSPs with high mass ($\gaeq 0.1M_\odot$) non-degenerate companions. One classic example is PSR J1023+0038 that has a $0.2M_\odot$ companion (Archibald et al. 2009). The X-ray/gamma-ray properties of \ufo\ are similar to that of PSR J1023+0038 but no radio emission from the former has been detected yet.

Hence, we suggest that \ufo\ is a new class of black widow-like MSP system, with no or very faint radio emission along the line-of-sight. The high-mass non-degenerate companion indicates that \ufo\ is in the late stages of recycling. The absence of radio emission is the result of different emission regions for radio and gamma-rays. For instance, the radio emission is from the polar cap with the radio beam out of the line-of-sight. On the other hand, the gamma-ray emission is from the outer magnetosphere which is predicted by the outer gap model.

An intensive search for radio emission and pulsation from \ufo\ in the future can confirm the above scenario. At the same time, a search for gamma-ray/X-ray pulsation from the MSP will reveal the true nature of the system.

\begin{acknowledgements}
We thank the supporting staffs at the Lulin Observatory to arrange the service observations. The Lulin Observatory is operated by the Graduate Institute of Astronomy in National Central University, Taiwan. We also thank Roger Romani for providing insights and useful comments as well as his quick-look optical spectroscopic results. This project is supported by the National Science Council of the Republic of China (Taiwan) through grant NSC100-2628-M-007-002-MY3 and NSC100-2923-M-007-001-MY3. A.~K.~H.~K. gratefully acknowledges support from a Kenda Foundation Golden Jade Fellowship. C.~C.~C. and D.~D.~were supported in part by Chandra award GO0-11022A. C.~Y.~H. is supported by the National Research Foundation of Korea through grant 2011-0023383.
\end{acknowledgements}


\begin{references}
	
\reference{} Abdo, A.A.  et al. 2010a, ApJ, 715, 429
\reference{} Abdo, A.A. et al. 2010b, ApJS, 187, 460
\reference{} Abdo, A.A. et al. 2010c, ApJS, 188, 405
\reference{} Abdo, A.A. et al. 2011, ApJS, submitted, arXiv:1108.1435
\reference{} Archibald, A.M. et al. 2010, ApJ, 722, 88
\reference{}  Archibald, A.M. et al. 2009, Science, 324, 1411
\reference{} Arons, J. \& Tavani, M. 1993, ApJ, 403, 249
\reference{} Bogovalov, S.V., Khangulyan, D.V., Koldoba, A.V., Ustyugova, G.V., Aharonian, F.A. 2008, MNRAS, 387, 63 
\reference{} Camilo, F. et al. 2009, ApJ, 705, 1
\reference{} Caraveo, P.A. 2010, arXiv:1009.2421
\reference{} Cognard, I. 2011, ApJ, 732, 47
\reference{} Green, P.J. et al. 2004, ApJS, 150, 43
\reference{} Keith, M.J. et al. 2011, MNRAS, in press, arXiv:1109.4193
\reference{} Kotani, T., Kawai, N.,
Yanagisawa, K., et al.\ 2005, Nuovo Cimento C Geophysics Space Physics C,
28, 755
\reference{} Kulkarni, S.R., Phinney, E.S., Evans, C.R., Hasinger, G. 1992, Nature, 359, 300
\reference{} Laird, E.S. et al. 2009, ApJS, 180, 102
\reference{} Monet, D.G. et al. 2003, ApJ, 125, 984
\reference{} Muslimov, A.G. \& Harding, A.K. 2003, ApJ, 588, 430
\reference{} Pletsch, H.J. et al. 2011, ApJ, in press, arXiv:1111.0523
\reference{} Ransom, S.M. et al. 2011, ApJ, 727, L16
\reference{} Richards, G.T. et al. 2002, AJ, 123, 2945
\reference{} Romani, R.W. \& Shaw, M.S. 2011, ApJ, 743, L26
\reference{} Ruderman, M.A. \& Sutherland, P.G.  1975, ApJ, 196, 51
\reference{} Saz Parkinson, P.M. et al. 2010, ApJ, 725, 571 
\reference{} Stappers, B.W., Gaensler, B.M., Kaspi, V.M., van der Klis, M. \& Lewin, W.H.G. 2003, Science, 299, 1372
\reference{} Takata, J., Cheng, K.S. \& Taam, R.E. 2010, ApJ, 723, L68
\reference{} Takata, J., Cheng, K.S. \& Taam, R.E. 2011, ApJ, in press (arXiv:1111.3451)
\reference{} Takata, J., Wang, Y. \& Cheng, K.S. 2010, ApJ, 715, 1318
\reference{} Takata, J., Wang, Y. \& Cheng, K.S. 2011, ApJ, 726, 44
\reference{} Tam, P.H.T., Hui, C.Y., Huang, R.H.H., Kong, A.K.H., Takata, J. et al. 2010, ApJ, 724, L207
\reference{} Tam, P.H.T., Huang, R.H.H., Takata, J., Hui, C.Y., Kong, A.K.H., Cheng, K.S. 2011, ApJ, 736, L10
\reference{} Venter, C. Harding, A.K. \& Guillemot, L. 2009, ApJ, 707, 800
\reference{} Wilson, C.A., Dieters, S., Finger, M.H., Scott, D.M., van Paradijs, J. 1999, ApJ, 513, 464

\end{references}
\end{document}